%% file: ddm_arXivresubmit.tex
\documentclass[aps,prd,showpacs,preprintnumbers,nofootinbib,twocolumn]{revtex4}  %

\usepackage{epsfig}
\usepackage{amsmath}
\usepackage{amssymb}

\begin{document}


\input{mymacros.tex}     
\title{
Lifetime Constraints for Late Dark Matter Decay
}
\author{Nicole F. Bell, Ahmad J. Galea and Kalliopi Petraki}
\affiliation{School of Physics, The University of Melbourne, Victoria 3010, Australia}


\begin{abstract}
We consider a class of late-decaying dark-matter models, in which a
dark matter particle decays to a heavy stable daughter of
approximately the same mass, together with one or more relativistic
particles which carry away only a small fraction of the parent rest
mass.  Such decays can affect galactic halo structure and evolution,
and have been invoked as a remedy to some of the small-scale
structure-formation problems of cold dark matter.
There are existing stringent limits on the dark matter lifetime if
the decays produce photons.
By considering examples in which the relativistic decay products instead
consist of neutrinos or electron-position pairs, we derive
stringent limits on these scenarios for a wide range of dark matter
masses.
We thus eliminate a sizable portion of the parameter space for these
late-decay models if the dominant decay channel involves Standard
Model final states.

\end{abstract}


\maketitle


\section{Introduction}

Despite the ample gravitational evidence for the existence of dark
matter (DM), its nature remains unknown. A promising strategy for
identifying the dark-matter particle is to search for a signature
produced by DM decay or annihilation, inside the Galactic halo or at
cosmological distances. Although such a signature has not yet been
identified, the observed radiation backgrounds have served to
constrain various decay and annihilation
channels~\cite{Yuksel:2007dr,PalomaresRuiz:2007ry,PalomaresRuiz:2008ef,PalomaresRuiz:2007eu,Zhang:2009pr,Zhang:2009ut,Beacom:2006tt,Yuksel:2007ac,Bell:2008vx,Cirelli:2009dv,Collaboration:2010rg,Mack:2008wu,Crocker:2010gy}. In many cases, this has resulted in stringent limits on specific DM
candidates.

In this paper we derive constraints for a class of models in which DM
decays dominantly via the process
\beq \x \rightarrow \x' + l,
\label{decay channel}
\eeq
where $\x'$ is a massive stable particle, and $l$ denotes one or more
light (or possibly massless) particles.
Such decay modes have been discussed in
Refs.~\cite{Cembranos:2005us,Kaplinghat:2005sy,SanchezSalcedo:2003pb,Abdelqader:2008wa,Peter:2010au,Peter:2010jy}.
We focus on the case where the mass splitting between the
unstable parent $\x$ and the daughter
particle $\x'$ is small,
\begin{equation}
\D m = m_\x - m_{\x'} \equiv \ve m_\x,
\end{equation}
where $\ve \ll 1$.  Since the decay replaces an $\x$ DM particle with
an $\x'$ DM particle of approximately the same mass, it does not
change the DM energy density significantly and it is thus not
constrained to occur at time scales much larger than the age of the
Universe.  This is quite different from decaying DM scenarios in which
the entire energy of the DM is converted into relativistic species
during the decay, such as the model of Ref.~\cite{Cen:2000xv}.  The
latter scenario is stringently constrained by the cosmic microwave
background anisotropies, which set a lower limit on DM lifetime of
123~Gyr at $1\s$ confidence level~\cite{Ichiki:2004vi}. The DM
decay models we consider in this paper, in which most of the energy is
retained in the form of nonrelativistic matter, may be most
effectively constrained by comparing the flux of the light particle(s)
$l$ emitted in the decay against the observed radiation backgrounds.

\begin{figure}[t]
\centering
\includegraphics[width=\linewidth]{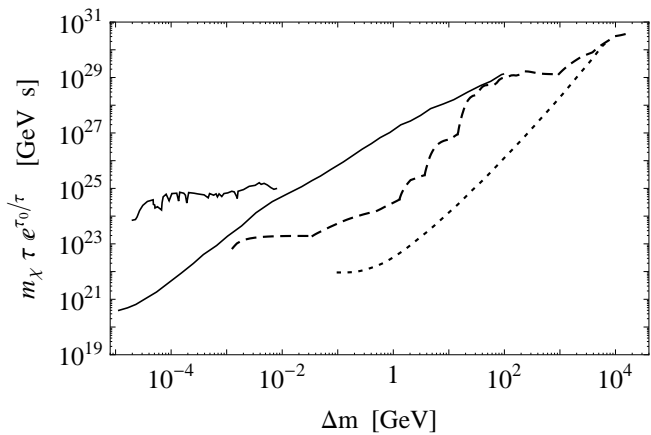}
\caption{
Constraints on the dark-matter decay channels:
(i) $\x \rightarrow \x' + \g$,  using the $\g$-ray line emission limits from
the Galactic center region and the isotropic diffuse photon background
(irregular and smooth solid lines, respectively)~\cite{Yuksel:2007dr};
(ii) $\x \rightarrow \x' + e^- + e^+$ (dashed line);
(iii) $\x \rightarrow \x' + \n + \bar{\n}$ (dotted line).
The regions below the lines are excluded.
}
\label{mtauVdm}
\end{figure}

An interesting feature of this class of models is that they offer a
possible solution to some of the discrepancies between galactic
observations and small-scale structure predictions of cold dark matter
(CDM) simulations. In CDM simulations, clustering proceeds
hierarchically, in a ``bottom-up" fashion, resulting in rich structure
at small scales. Several disparities with galactic observations have been identified~\cite{Gentile:2004tb,Salucci:2007tm,Diemand:2006ik,Gilmore:2006iy,Gilmore:2007fy,Gilmore:2008yp,Wyse:2007zw,Governato:2002cv,Klypin:1999uc,Metcalfe:2000nv,Moore:1999nt,Peebles:2001nv,SommerLarsen:1999jx,Tikhonov:2009jq}. For the disagreement to be resolved within the CDM paradigm, some
mechanism that modifies the standard CDM structure-formation picture
is necessary. If dark matter decays according to Eq.~(\ref{decay channel})
at time scales comparable to the age of the Universe, the
kinetic energy acquired by the heavy daughter, $\x'$, will effectively
heat up the dark-matter haloes, and cause them to expand. This can
alleviate two of the most glaring CDM problems, the cuspy density
profiles of dark-matter haloes and the over-prediction of satellite
galaxies~\cite{Cembranos:2005us,Kaplinghat:2005sy,SanchezSalcedo:2003pb,Abdelqader:2008wa,Peter:2010au,Peter:2010jy}.

Y\"{u}ksel and Kistler~\cite{Yuksel:2007dr} obtained bounds for the
decay channel $\x \rightarrow \x' + \g$, using observations by SPI,
COMPTEL and EGRET.  Since photons are the most easily detectable
particles of the Standard Model (SM), these bounds constrain
late-decaying DM scenarios in the most stringent fashion.  However,
the branching ratio for direct photon production may be tiny in many
models.
In this paper we investigate DM decay modes to other possible SM
particles.  We focus on two cases: when the light daughter particles
$l$ produced in the decay are a pair of the lightest charged leptons
of the SM
\beq
\x \rightarrow \x' + e^- + e^+,
\label{e channel}
\eeq
or a pair of the least detectable stable SM particles
\beq
\x \rightarrow \x' + \n + \bar{\n}.
\label{nu channel}
\eeq
Given that photons provide the most stringent bounds, and neutrinos
the least stringent, our results, together with those
of~\cite{Yuksel:2007dr}, span the full range of constraints for all SM
final states.  Moreover, constraining the decay into neutrinos places
a general lower bound on the DM lifetime for decay to \it{any} SM
particle.

The production of electrons and positrons by dark-matter decay can be
constrained both by the observed positron flux and by the observed
photon background. Electrons and positrons propagate in the Galaxy and
produce photons in a variety of ways. We consider a comprehensive list
of effects which result in photon emission associated with the
dark-matter decay into $e^\pm$ pairs, including inverse Compton
scattering, bremsstrahlung, synchrotron radiation, in-flight and
at-rest annihilations and internal bremsstrahlung. We compare the
expected signals with data from Fermi LAT, COMPTEL, EGRET and
INTEGRAL.  The decay into $\nu \bar{\nu}$ can be constrained by the
atmospheric-neutrino background, and by data from the Super-Kamiokande
search for the diffuse supernova neutrino background.

In Sec.~\ref{constraints} we derive the constraints on the decay
channels (\ref{e channel}) and (\ref{nu channel}). We present our
results in Fig.~\ref{mtauVdm}, together with the constraints derived
in Ref.~\cite{Yuksel:2007dr} for decay to a photon.
In Sec.~\ref{SSS}, we investigate whether these limits permit DM decay
to have a significant effect on the formation of galactic structure.
We focus on late-decay scenarios, at times $\gtrsim 0.1 \Gyr$, in
accordance with the work of
Refs.~\cite{Abdelqader:2008wa,SanchezSalcedo:2003pb,Peter:2010au,Peter:2010jy}.
For the decay modes described in Eq.~(\ref{e channel}), and
Eq.~(\ref{nu channel}), our constraints eliminate a sizable band of
the $\t$ vs $m_\x$ parameter space.  However, significant parameter
space for which DM decay may influence halo structure remains
available.
In Sec.~\ref{models} we turn our attention to models which are not 
necessarily motivated by the structure formation debate, nevertheless 
they involve decay of a relic population of particles into states of 
similar mass. In such models, the decaying particles may amount only 
to a small fraction of the total DM density of the Universe. We explore 
the applicability of our constraints on these scenarios and comment 
on specific particle-physics models.


\section{Constraints on dark-matter decay \label{constraints}}

The energy of the light particles produced in the decays depends on
the mass splitting between the parent and the heavy daughter particle,
$\D m$.
Their flux is inversely proportional to the lifetime, $\t$, and the mass, $m_\x$,
of the parent particle. The latter determines the number density of the DM particles.
Observations thus set a lower limit on the product $m_\x \t$ as a function of $\D m$.

However, if $\t$ is comparable to the age of the Universe $\t_0 \sim 13 \Gyr$,
an essential element in the proposed late-decaying DM resolutions of the
small-scale structure problems of
CDM~\cite{Abdelqader:2008wa,SanchezSalcedo:2003pb,Peter:2010au,Peter:2010jy},
the abundance of $\x$ particles may have changed significantly due to decay.
Observational bounds should then be cast in the form $m_\x \t \: e^{\t_0/\t} \geqslant \mathcal{F} (\D m)$,
reflecting the exponential suppression of the parent-particle number density today.
(The observational bound, $\mathcal{F} (\D m)$, is presented in
Fig.~\ref{mtauVdm} and will be discussed further below.)
Note that for sufficiently short lifetime, $\t \ll \t_0$, all the $\x$
particles decay to the stable $\x'$ state very early on, thus leaving no
observational signal in the Universe today, while for sufficiently
long lifetimes, $\t \gg \t_0$, the current decay probability will
be very small.
The lifetimes of interest to us, i.e., those that we can probe via
indirect detection of the DM decay products, lie between these two
extremes.

In our analysis below, we will assume that the $\x$ particles
originally constitute all of the DM, with no significant admixture of
the similar-mass $\x'$ particles.
Comparable initial abundances of $\x$ and $\x'$ might be anticipated
in canonical thermal-relic realizations of such a DM model.  However,
our limits can easily be rescaled to account for multicomponent DM,
and a down-scaling of our constraints by a factor of $\sim 2$ would
not qualitatively affect our conclusions.

The massive daughter particle $\x'$ will be produced
nonrelativistically, provided that $\D m \ll m_\x$.  Then, to a good
approximation, the light particles share the energy $\D m$ released in
the decay, such that $E_1 + E_2 \simeq \D m$, where $E_1,E_2$ are the
energies of the two light leptons.  For the 3-body decay channels
described in Eq.(\ref{e channel}) and Eq.(\ref{nu channel}), the
energy spectra of the daughter particles is in general
model-dependent.  However, we find that the constraints on the
dark-matter decay are rather insensitive to the actual spectrum.  We
illustrate this by presenting results for two limiting cases of the
daughter energy spectra:
\begin{enumerate}
\renewcommand{\theenumi}{\roman{enumi}}
\item Monoenergetic, where the energy spectrum per decay for
each light daughter species is
\beq
dN/dE = \d (E - \D m/2).
\eeq
\item
Flat, where is spectrum is uniform over the allowed energy range
such that
\beq
dN/dE = 1/(\D m-2m_l),
\eeq
where $m_l=m_e$ for the decay into
$e^\pm$ and $m_l=0$ for the decay into neutrinos.
\end{enumerate}
For each of the decay
modes analyzed, we pick the most conservative limits to further
continue our discussion\footnote{As a consistency check, we verify that
for a specific $\D m$, the limit obtained on $m_\x \t \: e^{\t_0/\t}$
when an extended injection distribution is used, does not exceed the
maximum value of this limit obtained for monoenergetic
injection for mass splittings up to $2 \D m$, i.e.
$\sqpare{m_\x \t  \: e^{\t_0/\t} (\D m)}\ssub{ext} \leqslant \max_{_{\m <2 \D m}} \sqpare{m_\x \t \: e^{\t_0/\t} (\m)}\ssub{mono}$}.

The constraints on DM decay depend on the DM halo density profile.
Although there are uncertainties in the profile, particularly near the
Galactic center (GC), we shall see below that these uncertainties are
subdominant.  For consistency with the results of other analyses which
we shall employ in our calculations, we use the Navarro-Frenk-White
(NFW)~\cite{Navarro:1995iw} profile throughout. The standard
parametrization is
\beq
\r (r) = \frac{\r_0}{(r/r_s)^\g [1+ (r/r_s)^\a]^{(\b-\g)/\a}},
\label{NFW}
\eeq
where $(\a,\b,\g) = (1,3,1)$. For the Milky Way $r_s = 20 \kpc$, and
the normalization $\r_0$ is fixed such that at the solar-circle
distance $R\sub{sc} = 8.5 \kpc$, the density is $\r(R\sub{sc}) = 0.3
\GeV \cm^{-3}$.

\subsection{Decay into electrons and positrons.}
\begin{figure}[t]
\centering
\includegraphics[width=\linewidth]{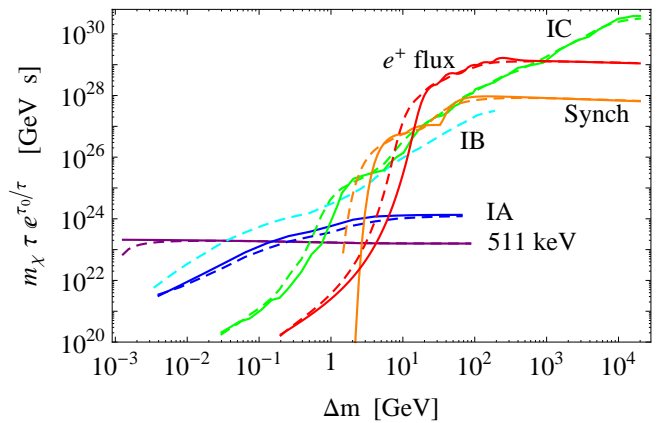}
\caption{
Limits on dark-matter decay process $\x \rightarrow \x' + e^- + e^+$, obtained from
internal bremsstrahlung (IB, cyan),
in-flight annihilation (IA, blue),
annihilation at rest (511 keV, purple),
inverse Compton scattering and bremsstrahlung emission (IC, green),
synchrotron radiation (Synch, orange),
and positron flux ($e^+$ flux, red).
Solid lines correspond to a monoenergetic spectrum of electrons and positrons,
while dashed lines assume a flat injection spectrum over the allowed energy range.
The regions below the lines are excluded.}
\label{mtauVdm e+e-}
\end{figure}

Electrons and positrons injected in the Galaxy will produce photon
fluxes spanning a wide range of energies, by a variety of different
mechanisms.  The injection rate can thus be constrained by the
resulting contribution to the Galactic photon backgrounds.  We will
also derive constraints based upon direct detection of positrons.
Below, we describe the methods used to estimate the signal produced by
each effect.  We present the resulting constraints in
Fig.~\ref{mtauVdm e+e-}.

\subsubsection{Internal bremsstrahlung.\label{SecIB}}
Electromagnetic radiative corrections to the DM decay channel to
charged particles inevitably produce real $\g$ rays. If the process
$\x \rightarrow \x' + e^\pm$ takes place, then $\x \rightarrow \x' +
e^\pm + \g$ must also take place since either one of the final state
charged particles can radiate a photon.  This process is called
internal bremsstrahlung (IB).  Note that the photon in the IB process
arises directly in the Feynman diagram for the decay, and is unrelated
to regular bremsstrahlung which results from interaction in a medium.
Importantly, the IB spectrum and normalization is approximately 
model-independent.  In addition, as the IB photon flux does not depend on the
propagation of charged decay products in the Galaxy, it is free of
uncertainties associated with the astrophysical environment.  IB is
thus a very clean and reliable technique for obtaining limits on DM
decay to charged particles, with the only source of uncertainty
residing in the DM profile adopted.

The IB differential decay width is~\cite{Beacom:2004pe}
\beq
\frac{d \: \G\sub{IB}}{dE} = \G \: \times h(E),
\label{IB decay rate}
\eeq
where $E$ is the energy of the photon emitted, $\G = \t^{-1}$ is the
rate of lowest order decay process $\x \rightarrow \x' + e^\pm$, and
$h(E)$ is the photon spectrum per $\x \rightarrow \x' + e^\pm$ decay.
This spectrum is independent of the new physics which mediates the
decay, and is given by
\beq
h(E) \simeq 
\frac{\a}{\p} \: \frac{1}{E} \ln \pare{\frac{s'}{m_e^2}}  \sqpare{1 + \pare{ \frac{s'}{s}}^2},
\label{IB h(E)}
\eeq
where $s' = s(1-E/E\sub{max})$, with $s$ being the energy released in
the decay, and $E\sub{max}$ the maximum energy that can be imparted to
the radiated photon\footnote{These general definitions of $s, \: s'$ reproduce the correct expressions for the case of annihilation into two charged particles~\cite{Beacom:2004pe} and for muon decay~\cite{Mardon:2009rc}.}.
In the 3-body decay we consider here, $s = (\D m-2m_e)^2$, and
$E\sub{max} = \sqrt{s} (1-\ve/2) \simeq \sqrt{s}$.\footnote{Note that the phase space available to the photon when one of the decay products is emitted nonrelativistically (i.e. when $\ve \ll 1$) is larger than when all decay products are relativistic (e.g. muon decay, where $\ve \simeq 1$). This is because a heavy daughter particle can absorb recoil momentum, without carrying away a significant amount of energy.}

The shape of the IB photon spectrum is determined by the $1/E$ factor
in Eq.~(\ref{IB h(E)}), while the logarithmic factor provides a sharp
cutoff at $E \simeq E\sub{max}$.
Since the galactic photon background varies approximately as $E^{-2}$,
the high-energy region of the IB photon spectrum will provide the most
stringent constraints.

\medskip

The flux of photons due to IB will be proportional to the dark matter density integrated over the line of sight.
At an angle $\ps$ from the Galactic center, the line-of-sight integral is
\beq
J(\ps) = \int_0^{l\sub{max}} dl \:\: \r_\x\pare{\sqrt{R\sub{sc}^2 - 2 l R\sub{sc} \cos \ps +l^2}},
\label{J}
\eeq
where $l\sub{max} = \sqrt{R\ssub{MW}^2 - R\sub{sc}^2\sin^2 \ps} +
R\sub{sc} \sin \ps$.  The average of $J$ over a cone of half-angle
$\ps$ centered on the GC is
\beq
\mathcal{J}\ssub{\D\W} (\ps) = \frac{2 \p}{\D\W} \int_0^\ps J(\psi') \sin \ps' d\ps',
\label{J aver}
\eeq
where $\D \W = 2 \p (1-\cos \ps)$.

\medskip

The expected photon flux, per unit solid angle, due to IB is then
\beq
\frac{d\F\ssub{IB}}{dE} = \frac{1}{4 \pi} \frac{1}{m_\x \t e^{\t_0/\t}} \mathcal{J}\ssub{\D\W} \: h(E).
\eeq

COMPTEL observations show that the $\g$-ray background remains approximately constant within $|l|<30^o$. Thus, along the Galactic plane, the signal-to-background ratio for IB is maximized at the GC. Here, we will average over the region $\ps<5^o$, which exceeds the angular resolution with which both COMPTEL and EGRET data have been reported~\cite{Strong:1998ck,Hunger:1997we}.
The IB limits are shown in Fig.~\ref{mtauVdm e+e-} (cyan line).

\subsubsection{In-flight and at-rest annihilations.\label{sec IA & 511}}

Positrons produced by DM decay will propagate in the interstellar
medium, lose energy, and annihilate with electrons.  Most positrons
will survive until they become nonrelativistic, and annihilate at
rest to produce a 511 keV annihilation line (together with a three-photon
continuum).  However, a portion of the positrons annihilate in flight
while still relativistic.  These in-flight annihilations (IA) can
produce a significant flux of $\g$ rays~\cite{Beacom:2005qv}.

Reference~\cite{Beacom:2005qv} considered the injection of positrons at the
Galactic center, and related the flux of IA $\g$-rays to the flux of
511 keV photons.  This was used to derive a constraint on the
injection energy of positrons that contribute to the observed Galactic
511 keV line.

We will follow a similar approach to obtain limits on DM decay into
$e^\pm$. In our case, however, the overall normalization of the photon
flux is not fixed by the observed 511 keV line intensity, but is
instead determined by the DM decay rate. The continuum $\g$-ray flux
due to IA, and the 511 keV photons from annihilation at rest, thus
yield thus two separate sets of constraints.  The positron injection,
deceleration and annihilation budget is described by the equation~\cite{Strong:1998pw}
\beq
\frac{\partial n}{\partial t} + \frac{\partial}{\partial \g} \pare{\frac{d \g}{d t} n}  = -n\ssub{H} \s(\g) \beta \: n(\g,t) + \G\sub{inj} \: \frac{dN}{d\g},
\label{continuity}
\eeq
where $n(\g,t) d\g$ is the density of positrons with energy $\g m_e$,
at time $t$, $\G\sub{inj} = \r_\x e^{-\t_0/\t} / m_\x \t$ is the positron-density
injection rate due to DM decay, and $dN/d\g$ is the positron spectrum
per decay. The positron annihilation cross section $\s(\g)$ on
electrons at rest is~\cite{Dirac}
\beq
\s = \frac{\pi r_e^2}{\g + 1} \sqpare{ \frac{\g^2 + 4\g +1}{\g^2-1} \ln \pare{\g + \sqrt{\g ^2-1}} - \frac{\g  + 3}{\sqrt{\g^2-1}} },
\label{sigma}
\eeq
where $r_e$ is the classical electron radius.

At energies lower than $\sim$ 1 GeV, ionization and Coulomb losses (in
a neutral or an ionized medium, respectively) dominate over energy
loss due to synchrotron, bremsstrahlung and inverse Compton
effects, while at higher energies synchrotron losses are more
important.  The energy loss rates due to ionization, Coulomb
scattering and synchrotron emission are~\cite{Strong:1998pw}
\bea
\left| \frac{d \g}{dt} \right|\sub{i}
&\simeq&  4.4 \cdot 10^{-15} \: \rm{s}^{-1} \: \left(\frac{n\ssub{H}}{0.1 \cm^{-3}}\right) \frac{\ln \g + 6.8}{\b} \label{dg/dt -ion},
\label{eq:i}
\\
\left| \frac{d \g}{dt} \right|\sub{C}
&\simeq&  1.5 \cdot 10^{-15} \: \rm{s}^{-1} \: \left(\frac{n\ssub{H}}{0.1 \cm^{-3}}\right) \frac{\ln \g + 75.9}{\b} \label{dg/dt -c},
\label{eq:C}
\\
\left| \frac{d \g}{dt} \right|\sub{s}
&\simeq&  5.4 \cdot 10^{-24} \: \rm{s}^{-1} \: \left(\frac{U_B}{0.2 \: \frac{\eV}{\cm^3}}\right) (\g^2 -1) \label{dg/dt -syn},
\eea
where $n\ssub{H}$ is the hydrogen number density of the medium, and
$U_B=B^2/8\p$ is the energy density of the magnetic field.
Equations~(\ref{eq:i}) and Eq.(\ref{eq:C}) apply to a fully neutral and or
fully ionized medium, respectively.  As a conservative choice, we set
the ionization fraction of the interstellar gas to be $x_i \approx
0.51$.  Compared with a completely neutral medium, energy losses in an
ionized medium are larger and thus IA constraints weaker, by a factor
of a few~\cite{Sizun:2006uh}

As evident from Eqs.~(\ref{dg/dt -ion}) and (\ref{dg/dt -c}), it takes
approximately $10^7 \yr$ for relativistic positrons to slow down and
annihilate. If positron injection (DM decay) has been taking place
over a much larger time period, $\t > 0.1 \Gyr$, the positrons reach a
steady-state distribution. The time-independent solution of
Eq.(\ref{continuity}) is
\beq
n(\g) = \frac{\r_\x}{m_\x \t \: e^{\t_0/\t}} \: \frac{1}{\abs{\frac{d \g}{d t}}}
\int_\g^\infty d\g' \: \frac{d N}{d\g'} P_{\g' \rightarrow \g},
\label{n}
\eeq
where
\beq
P_{\g' \rightarrow \g} \equiv \exp \left[-n\ssub{H} \int_{\g}^{\g'}\frac{\sigma(\g'') \b''}{\left|\frac{d \g''}{d t}\right|}d\g'' \right]
\label{surv prob}
\eeq
is the \it{survival probability}, introduced in \cite{Beacom:2005qv},
of positrons injected at energy $\g' m_e$ to have not annihilated by
the time they reach energy $\g m_e$.

Here, we have ignored the effects of spatial diffusion of positrons,
which would require detailed modeling. This approximation is
sufficient provided that the positrons remain trapped within the
region we consider\footnote{Along with spatial diffusion, we also
  ignore the related effect of reacceleration. This becomes more
  important with increasing energy, since the diffusion coefficient in
  momentum space is $D_{pp} \propto p^{2-\d}$ with
  $\d<1$~\cite{Strong:1998pw,Zhang:2009pr,Zhang:2009ut}. However, the
  probability to annihilate drops sharply at high energies, and thus
  pumping more energy in already energetic positrons does not affect
  the IA constraints significantly.}.
In what follows, we will calculate the expected photon flux due to
annihilation in a cylindrical region of radius $r\sub{max}=3 \kpc$ and
half-height $z\sub{max}=0.5 \kpc$, centered at the GC. This is well
inside what is taken to be the diffusion area in most models
simulating cosmic-ray propagation in the Galaxy. Because of the Galactic
magnetic field, charged particles are largely expected to follow the
magnetic lines and converge toward the GC, as long as their Larmor
radius, $r_L = \g m/q B$, is much smaller than the scale of variation
of the magnetic field. In the Milky Way, $B \gtrsim \rm{few} \: \m
\rm{G}$, and $(d \ln B/dr)^{-1} \sim$~few~kpc~\cite{Zhang:2009pr}.
Positrons are thus expected to follow the magnetic lines, even for
energies much higher than the ones considered in this analysis.  Any
influx of positrons in the region under consideration would only lead
to more stringent constraints than the ones derived here.  (The
possibility of transport of the positrons from the disk to the GC has
been discussed in detail in \cite{Prantzos:2005pz}.)

The diffuse $\g$-ray background of the Galactic plane has been
measured by COMPTEL in the energy range 1--30 MeV for the region
$|l|<30^o$, $|b|<5^o$ (Galactic longitude and latitude, respectively)
and by EGRET, in the energy range 30 MeV--30 GeV and the region
$|b|<10^o$~\cite{Strong:1998ck,Hunger:1997we}.  The cylindrical volume
defined above is covered by these observations, and encompassed within
a conical patch of solid angle $\D \W = 4 l\sub{max} \sin b\sub{max}
\simeq 0.13 \sr$, with $l\sub{max} = \arcsin (r\sub{max}/R\sub{sc})
\simeq 20^o$ and $b\sub{max}= \arctan
       [z\sub{max}/(R\sub{sc}-r\sub{max})] \simeq 5^o$.  The
       comparison of the signal from the cylindrical region with the
       background radiation emanating from the larger conical region,
       keeps our constraints conservative.

\bigskip

The flux of photons (per unit solid angle) due to IA is
\[
\frac{d \F\sub{IA}}{d k} = \frac{1}{\D \W} \: \frac{1}{4 \p R^2\sub{sc}} \: \int dV \: \int d\g \: n(\g) n\ssub{H}  \frac{d \sigma(k,\g)}{d k} \b,
\]
where the spatial integration is over the cylindrical region
considered. The differential cross section $\frac{d \sigma(k,\g)}{d k}$
for a positron of energy $\g m_e$ to produce a photon of energy
$E = k m_e$ is~\cite{Beacom:2005qv,Stecker:1969}
\beq
\frac{d \s(k,\g)}{d k} = \frac{\pi r_e^2}{\g^2-1} \left[\frac{-\frac{3+\g}{1+\g} + \frac{3+\g}{k} - \frac{1}{k^2}}{\left(1-\frac{k}{1+\g}\right)^2}-2\right],
\label{sigma-diff}
\eeq
while the minimum positron energy $\g\sub{min} m_e$ required to produce a photon of energy $k m_e$ is given by
\beq
\g\sub{min}(k) = k-\frac{1}{2} + \frac{1}{2 (2 k-1)}.
\eeq
For the steady-state positron distribution $n(\g)$, given in
Eq.(\ref{n}), the expected $\g$-ray signal from IA is
\beq
\frac{d \F\sub{IA}}{d k} = \frac{1}{ m_\x \t \: e^{\t_0/\t}} \: \frac{\int dV \: \r_\x}{4 \p R\sub{sc}^2 \: \D \W} \: \int_{\g\sub{min}(k)}^\infty d\g \: \frac{dN}{d\g} \: \mathcal{Q}(k,\g),
\label{flux-diff}
\eeq
where
\beq
\mathcal{Q}(k, \g) \equiv n\ssub{H}  \:
\int_{\g\sub{min}(k)}^\g  \!\!\!\!  d\g' \: \frac{\sqrt{\g'^2-1}}{\g'} \: \frac{d\s(k,\g')}{dk}  \frac{P_{\g \rightarrow \g'}}{\abs{\frac{d\g'}{dt}}}.
\label{Q}
\eeq

The IA flux of Eq.~(\ref{flux-diff}) is compared to the $\g$-ray
background reported by COMPTEL and
EGRET~\cite{Strong:1998ck,Hunger:1997we}, and compiled in
Ref.~\cite{Mack:2008wu}.  The resulting lower limit on $m_\x \t \: e^{\t_0/\t}$ is
presented in Fig.~\ref{mtauVdm e+e-} (blue line) and increases with
increasing $\D m$, since positrons injected at higher energy spend
more time as relativistic, and are more likely to annihilate in
flight. However, the annihilation probability diminishes at high
energies, which is manifest by the plateau in the lifetime bound for
energies $\gtrsim 3 \GeV$.

\bigskip

The majority of positrons survive until they become
nonrelativistic. Then, they either annihilate directly with electrons
to produce 511 keV photons, or they form a positronium bound
state. Positronium subsequently annihilates, with probability 25\%
into two 511 keV photons, or with probability 75\% into a three photon
continuum.  The positronium fraction in the Galaxy is fixed by the
relative intensities of the $\g$-ray continuum below 511 keV, and the
511 keV line. In our Galaxy $f = 0.967 \pm 0.022$~\cite{Jean:2005af}.

If DM decays into $e^\pm$ pairs, the flux of 511 keV photons produced
is determined, in the steady-state regime, by the rate at which
positrons arrive at rest. For each nonrelativistic positron, there
will be $2(1-f) + 2 f/4 = 2(1-3f/4)$ photons contributing to the
line. The expected flux of 511 keV photons is
\beq
\F_{511} =  \frac{2(1-3f/4)}{m_\x \t  \: e^{\t_0/\t}} \:  \frac{\int dV \: \r_\x}{4 \p R\sub{sc}^2 \: \D\W} \: \int_1^\infty d\g \: \frac{dN}{d\g}  P_{\g \rightarrow 1}.
\label{flux 511 keV}
\eeq

The $\g$-ray observations by INTEGRAL have revealed a 511~keV line
emission of intensity $0.94 \times 10^{-3} \: \rm{ph \: cm^{-2} \:
  s^{-1}}$ within an integration region $|l|<20^o$,
$|b|<5^o$~\cite{Knodlseder:2005yq}. The average flux per steradian is
then $0.0077 \: \rm{ph \: cm^{-2} \: s^{-1} \: sr^{-1}}$.  The limits
on DM decay arising from annihilation at rest are presented in
Fig.~\ref{mtauVdm e+e-} (purple line). They exhibit a very slight
negative slope towards increasing $\D m$, which accounts for the
(small) portion of positrons annihilating in-flight when injected at
high energies.

The IA constraints are stronger than those for annihilation at rest
for average positron injection energies $\langle E \rangle = \D m/2
\gtrsim 80 \MeV$.  This is considerably higher than the corresponding
value of $\sim 3 \MeV$ found by an appropriate analysis in
Ref.~\cite{Beacom:2005qv}.  The latter limit was determined by
requiring that the IA $\g$-rays do not exceed 30\% of the observed
background, instead of the more conservative 100\% adopted here.
However, the more important factor in reconciling these two analyses
is the different signal regions considered.  The analysis in
Ref.~\cite{Beacom:2005qv} considered the conical region within a
$\sim 0.37 \kpc$ radius of the GC, which contains the peak
of the 511 keV annihilation signal.  This is to be compared to our larger
cylindrical volume ($r\sub{max}=3 \kpc$, $z\sub{max}=0.5 \kpc$) outlined above.
The effect of choosing the larger volume is to make the
annihilation-at-rest constraints stronger, with respect to those from
IA, than for the choice of a smaller region at the Galactic center.
The Galactic center is not the optimal observation region to use in
setting an annihilation-at-rest limit, because that is where the
observed 511 keV background is highest.  In addition to resulting in
more sensitive annihilation-at-rest limits, the larger volume also
renders us insensitive to possible effects of diffusion.

Notice also that the photon fluxes, Eqs.~(\ref{flux-diff}) and
(\ref{flux 511 keV}), are independent of the medium number density
$n\ssub{H}$, as long as ionization and Coulomb losses dominate, for
$\D m \lesssim \rm{few} \GeV$. Since the probability to annihilate at
even higher energies is insignificant, the results are rather
insensitive to $n\ssub{H}$, as well as $U_B$, throughout the energy
spectrum.

\subsubsection{Inverse Compton scattering, bremsstrahlung emission, synchrotron radiation.\label{SecIC-B-Syn}}

At high energies IA become rather rare, and electrons and positrons
propagating in the Galaxy produce radiation more efficiently via other
mechanisms.  Gamma rays are produced by bremsstrahlung and inverse
Compton (IC) scattering of low-energy photons, with the latter effect
yielding the dominant contribution. Synchrotron emission in the
Galactic magnetic field gives rise to a radio wave signal.

Zhang et al.~\cite{Zhang:2009pr,Zhang:2009ut} calculated the photon
spectrum expected if electrons and positrons are injected in the
Galaxy by DM decay. They assumed a monoenergetic injection spectrum,
and modeled the propagation of $e^\pm$ in the interstellar medium in
detail, including the effects of spatial diffusion, convection, energy
loss and reacceleration. They compared the resulting spectrum at
Earth with observed backgrounds, and encoded their calculations in
\textit{response functions} which can be utilized to obtain
constraints when convoluted with the DM decay spectrum of a particular
model. Here, we use the response functions derived in
Ref.~\cite{Zhang:2009pr} for synchrotron radiation, and in
Ref.~\cite{Zhang:2009ut} for IC and bremsstrahlung emission, to
obtain constraints for the $\x \rightarrow \x' + e^\pm$ decay process.

The main source of uncertainties in deriving the response functions is
poor knowledge of the various astrophysical parameters which determine
the propagation of $e^\pm$ in the Galaxy. The most significant
contribution comes from the height of the diffusion zone. The Galactic
magnetic field, although also quite poorly known, contributes
subdominantly to the total uncertainty. This is true even for the
synchrotron response functions, since although the magnetic field may
be rather uncertain in the Galactic bulge, the directions which
optimize the signal-to-background ratio point away from the GC.
Different choices for the DM halo profile have a small influence on
the response functions, since the flux is only proportional to the
density (as opposed to the $\rho^2$ dependence applicable for DM
annihilation)~\cite{Zhang:2009pr,Zhang:2009ut}.

The IC and bremsstrahlung emission is constrained by using the Fermi
LAT $\g-$ray maps derived in~\cite{Dobler:2009xz}, for the energy
range 0.5~GeV--300~GeV. Because of reacceleration, this permits limits for
$e^\pm$ injection energies in a much wider range, $0.01 \GeV - 10^4
\GeV$. The strength of the limits, of course, sharply diminishes in
the low-energy part of this range.  At energies above 50 GeV, the
Fermi data may suffer from significant background contamination, while
point sources have not been subtracted. These effects lead to more
conservative constraints.  There are also various astrophysical
contributions to the $\g$-ray flux, such as nucleus-nucleus
photoproduction via $\p^0$ decay, and IC and bremsstrahlung radiation
from cosmic-ray electrons and positrons. In the spirit of setting
conservative limits on DM decay, we will use response functions where
these foregrounds have not been subtracted. The patch of sky chosen to
compare the expected signal to observations is $|l| < 20^o$ and $-18^o
< b < -10^o$~\cite{Zhang:2009ut}.

The synchrotron response functions derived in Ref.~\cite{Zhang:2009pr}
correspond to 408 MHz, 1.42 GHz and 23 GHz sky maps, and yield
constraints for $e^\pm$ injection energies in the range, $0.1 \GeV -
10^4 \GeV$. The directions that optimize the signal-to-background
ratio are different for each of the sky maps used, and are located
close to, but not at, the GC.
(Similar sensitivity may be achieved with the isotropic diffuse flux
measured by Fermi; see Ref.\cite{Chen:2009uq} for relevant limits on the
process $\chi \rightarrow e^\pm$.)

The response functions reported all assume a NFW halo profile. The
expected signal from particles produced by DM decay is compared to
observed photon flux including a 2$\s$ error. The dependence on the
propagation model, for both the synchrotron and the IC plus
bremsstrahlung emission, becomes more significant at $e^\pm$ injection
energies lower than about 10 GeV. This is because the limits in that
range rely solely on the effect of reacceleration, which depends
strongly on the model adopted, while no actual low-energy background
data are used.  For the synchrotron constraints, we adopt the ``DR''
model~\cite{Zhang:2009pr},
which yields moderately conservative constraints along the entire energy range.
For the IC plus bremsstrahlung constraints we adopt the ``L1'' model, whose response functions are most comprehensively reported in~\cite{Zhang:2009ut}.

The limits on DM decay from IC plus bremsstrahlung radiation and
synchrotron emission are shown in Fig.~\ref{mtauVdm e+e-} (green and
orange lines, respectively).

\subsubsection{Positron flux. \label{SecPosFlux}}

The DM decay into $e^\pm$ pairs can be constrained by direct
observations of the positron flux on Earth. Utilizing the same
propagation models as the ones used for the synchrotron emission,
Zhang et al.~\cite{Zhang:2009pr} constructed response functions,
comparing the expected positron flux to PAMELA
data~\cite{Adriani:2008zr}. The positron fraction reported by PAMELA
was converted into positron flux using the $e^\pm$ total flux observed
by the Fermi telescope~\cite{Abdo:2009zk}.

The positron-flux response functions are based on observations of
positrons in seven energy bands in the range $10.17 \GeV - 82.55
\GeV$. Depending on the model used, they can provide meaningful
constraints for positron injection energies within the range 
$0.1 \GeV - 10 \TeV$. These constraints appear to prevail over the IB, IA, IC
and synchrotron constraints at the high-energy part of the
spectrum. In the same spirit as before, we adopt the DR model, and
obtain fairly conservative constraints for the entire energy
spectrum. The results are presented in Fig.~\ref{mtauVdm e+e-} (red
line).

Much like the photon-emission response functions, the positron-flux
response functions are sensitive to the adopted propagation model,
which determines the diffusion and reacceleration of the injected
positrons. However, they are insensitive to the DM halo profile, since
most of the positrons come from the local region, $\sim 1 \kpc$ from
the Sun~\cite{Zhang:2009pr}\footnote{This is consistent with our
  assumption in Sec.~\ref{sec IA & 511}, that the $e^\pm$ do not
  escape from the inner part of the galaxy.}.

\subsection{Decay into neutrinos. \label{SecNuNubar}}

Neutrinos are the least detectable stable SM particles, hence
constraints on DM decay (or annihilation) to neutrinos can be used to
set conservative but robust lower limits on the DM lifetime (or upper
limit on the annihilation cross section) to any SM final
state~\cite{Beacom:2006tt}.
Palomarez-Ruiz~\cite{PalomaresRuiz:2007ry,PalomaresRuiz:2008ef} obtained limits on the
decay channel $\x \rightarrow \n + \bar{\n}$.  Here we will adapt the
analysis of Refs.~\cite{PalomaresRuiz:2007eu,PalomaresRuiz:2007ry,PalomaresRuiz:2008ef}, to
constrain the decay channel $\x \rightarrow \x' + \n + \bar{\n}$.
This will set the most conservative limit on DM decay modes of the
form $\x \rightarrow \x' + l$, where $l$ is any SM final state.

The neutrino flux on Earth from DM decay in the Galactic halo is
\beq
\frac{d\F_\n}{dE_\n}=\frac{1}{4\p}\frac{1}{m_\x\t e^{\t_0/\t}}\cal{J}_{\D\W}\frac{2}{3}\frac{dN}{dE_\n}
\label{nu flux}
\eeq
The multiplicity of 2 accounts for the sum of $\n$ and $\bar{\n}$
produced in the decays. As in Refs.~\cite{PalomaresRuiz:2007ry,PalomaresRuiz:2008ef}, we
assume equal decay width to all three neutrino flavors, which accounts
for the factor of $1/3$ in Eq.~(\ref{nu flux}).  This is a reasonable
approximation, since neutrino flavor oscillations between the
production and detection points will considerably weaken any
preference toward a particular flavor.
For simplicity, and in order to minimize the uncertainty arising from
a particular choice of halo profile, we average the expected neutrino
signal over the whole sky, $\ps=180^\circ$. Directional information,
whenever available, is in general expected to lead to more stringent
limits~\cite{PalomaresRuiz:2007ry,PalomaresRuiz:2008ef,Yuksel:2007ac}.

\begin{figure}[t]
\centering
\includegraphics[width=\linewidth]{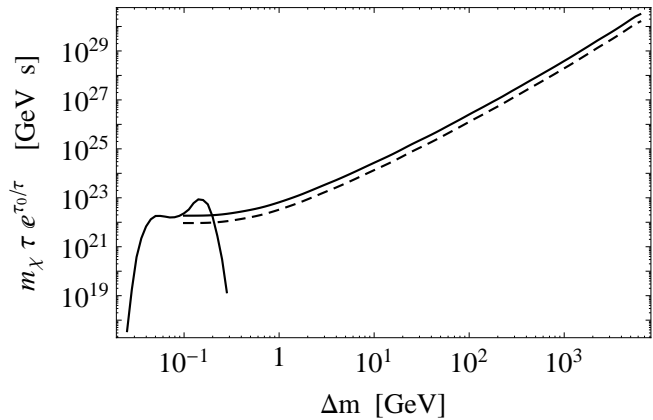}
\caption{Limits on dark-matter decay $\x \rightarrow \x' + \n + \bar{\n}$. As in Fig.~\ref{mtauVdm e+e-}, the solid (dashed) lines correspond to monoenergetic (flat) injection distribution of $\n, \: \bar{\n}$.  The regions below the lines are excluded.}
\label{mtauVdm nunubar}
\end{figure}

For energies $E_\n \gtrsim 50 \MeV$, the neutrino flux on Earth is
dominated by atmospheric neutrinos.  The atmospheric neutrino flux has
been well measured by a number of
experiments~\cite{Ashie:2005ik,Daum:1994bf,Ahrens:2002gq,:2007td,Ambrosio:2004ig,Allison:2005dt}
and is in good agreement with theoretical predictions.  We utilize the
results of the FLUKA~\cite{FLUKA,FLUKA2} calculation of the
atmospheric $\n_\m + \bar{\n}_\m$ background, over the energy range
50~MeV to 10~TeV.
We determine a DM decay limit by requiring the expected neutrino
signal from DM decay not to exceed the atmospheric neutrino flux,
integrated over energy bins of width $\D\log_{10}E_\n \sim 0.3$.  This
choice of bin size is in accordance with that adopted in
Refs.~\cite{PalomaresRuiz:2007ry,PalomaresRuiz:2008ef}, and
encompasses the experimental resolution of the neutrino detectors.
(We note that future measurement by IceCube+DeepCore will lead to
sensitive limits at high energy~\cite{Mandal:2009yk,Buckley:2009kw}.)

At low energies $E_\n \lesssim 100 \MeV$, the relevant data come from
the diffuse supernova neutrino background (DSNB) search performed by the
SK experiment~\cite{Malek:2002ns}.
SK searched for positrons produced by incident $\bar{\n}_e$ on free
protons inside the detector, via the inverse $\b$-decay reaction
$\bar{\n}_e + p \rightarrow e^+ + n$. Incoming $\n_e$ and $\bar{\n}_e$
interact also with bound nucleons, producing electrons and positrons.
The main sources of background for these observations are atmospheric
$\n_e$ and $\bar{\n}_e$, and the Michel electrons and positrons from
decays of subthreshold muons.  A DSNB signal was not detected.
In Refs.~\cite{PalomaresRuiz:2007eu,PalomaresRuiz:2007ry,PalomaresRuiz:2008ef} a $\x^2$
analysis of these data was performed, and used to place limits, at 90\% confidence level, on the contribution to the signal from DM annihilation, $\x\x
\rightarrow \n\bar{\n}$, or decay, $\x \rightarrow \n\bar{\n}$.
We shall employ the DM decay limits of
Refs.~\cite{PalomaresRuiz:2007ry,PalomaresRuiz:2008ef}.  A simple rescaling suffices to
convert these limits into constraints on the decay mode of interest
here, $\x \rightarrow \x' \n\bar{\n}$, under the assumption that
$\n\bar{\n}$ are emitted monoenergetically.
The resulting constraints are shown in Fig.~\ref{mtauVdm nunubar}.


\section{Late-decaying dark matter and the small-scale galactic structure \label{SSS}}

The standard $\L$CDM cosmological model successfully reproduces the
observed structure of the Universe at large scales.  However, at small
scales, there appear to be several inconsistencies between
observations and CDM simulations of galaxy formation.  The negligible
primordial velocity dispersion of CDM particles allows gravitational
clustering to occur down to very small scales. Structure forms
hierarchically, with small scales collapsing first and forming dense
clumps.  The resulting phase packing in the inner regions of galaxies
leads to cuspy central density profiles~\cite{Diemand:2006ik}, which
are currently disfavored by the rotational curves of dwarf spheroidal
galaxies~\cite{Gentile:2004tb,Salucci:2007tm,Gilmore:2006iy,Gilmore:2007fy,Gilmore:2008yp,Wyse:2007zw}.
Dense lumps of matter also survive the mergers and become satellite
galaxies. The number of satellite galaxies predicted by CDM
simulations greatly exceeds the number observed around the Milky Way
~\cite{Diemand:2006ik,Klypin:1999uc,Moore:1999nt}. A number of other
disparities between observations and CDM predictions have been
reported. These include the overestimation of the number of halos in
low-density voids~\cite{Peebles:2001nv,Tikhonov:2009jq}, the
non-prediction of pure-disk or disk-dominated
galaxies~\cite{Governato:2002cv}, the large angular-momentum loss by
condensing gas~\cite{SommerLarsen:1999jx}, and the ``bottom-up''
hierarchical formation~\cite{Metcalfe:2000nv}.

It is not yet known whether the solution to these inconsistencies lies
outside the dark-matter sector. Indeed, individual astrophysical
solutions to some of these problems have been suggested.  However, it
is possible that the apparent discrepancies between CDM and observed
galactic structure point toward a modification of the standard CDM
scenario.  The significance of these hints for deciphering the nature
of DM is underscored by the fact that gravitational effects remain the
only confirmed evidence for the existence of DM.

A possible alternative to the CDM model is warm dark matter (WDM),
whose thermal free-streaming properties result in a suppression of
structure on small scales.  Although well-motivated particle-physics
candidates for WDM exist, WDM lacks the ``naturalness" of the standard CDM
scenario in which the observed relic DM abundance arises from thermal
freeze-out (CDM candidates produced by other mechanisms also exist).
It is possible, though, to incorporate some WDM-like
structure-formation features within the CDM paradigm, if we discard
the notion that CDM particles are completely stable.

It has been suggested that if the heavy relic particles decay
according to Eq.(\ref{decay channel}), the energy released in the
decay will change the standard picture of structure formation.
If the decays occur at early times, $\sim 10^5 \: \rm{s} - 10^8 \:
\rm{s}$, before gravitational collapse, the massive decay products
acquire non-negligible velocities, which results in a suppression of the
amplitude of density perturbations at small scales (still in the
linear regime).  The daughter DM particles behave effectively as WDM.
Existing limits on the free-streaming and phase-packing properties of
WDM can be directly translated to determine the interesting and the
excluded regions of the CDM decay parameter
space~\cite{Cembranos:2005us,Kaplinghat:2005sy}.

Another interesting possibility, investigated in
Refs.~\cite{Abdelqader:2008wa,SanchezSalcedo:2003pb,Peter:2010au,Peter:2010jy},
arises if the decays take place during the nonlinear stages of gravitational collapse. In this
scenario the energy acquired by the decay products causes the halo to
expand and, as a result, softens the steep central cusps predicted by
CDM~\cite{SanchezSalcedo:2003pb}.
It also results in a decrease in the halo circular velocity, which can
account for the deficit of satellite galaxies with velocity 10-20
km/s, with respect to standard CDM
predictions~\cite{Abdelqader:2008wa}.

The efficacy and viability of the late DM decay mechanism to alleviate
the CDM small-scale structure problems depend on two parameters: the
parent particle lifetime, $\t$, and the ratio $\ve \equiv \D m/m_\x$
which gives the recoil velocity of the massive daughter particle,
$v_{\x'} \simeq \ve$.\footnote{This becomes a strict equality in the
case of a 2-body decay. If more than one relativistic particle is
produced, as in the cases examined in this paper, the correction in
$v_{\x'}$ is insignificant.}
Qualitatively, one expects $\t$ to be comparable to the age of the
Universe, so that the decays occur during galaxy formation.  In the
limit of very large $\t$, the standard CDM scenario is recovered,
while small $\t$ corresponds to the early decay scenario, examined
in~\cite{Cembranos:2005us,Kaplinghat:2005sy}.  The velocity imparted
to the heavy daughter particles should be sufficiently large -- of the
order of the virial velocity of the galaxy -- in order to have an
effect on the formation of the halo. It should not exceed, though, the
escape velocity from the halo, which would be catastrophic for the
galaxy, unless, of course, $\t$ is very large.

Abdelqader and Melia~\cite{Abdelqader:2008wa} used a semi-analytical
approach that allowed them to incorporate the effect of DM decay in
halo evolution. They showed that DM decay preferentially heats smaller
haloes, causing them to expand, and reduces their present-day circular
velocity. They argued that if $\ve \sim (5 - 7) \times 10^{-5}$ and
$\t \sim (\rm{few} - 30) \Gyr$, dark-matter decay can well account for
the deficit in the observed number of galaxies with circular
velocities in the range 10-20 km/s.

\begin{figure}[t]
\centering
\includegraphics[width=\linewidth]{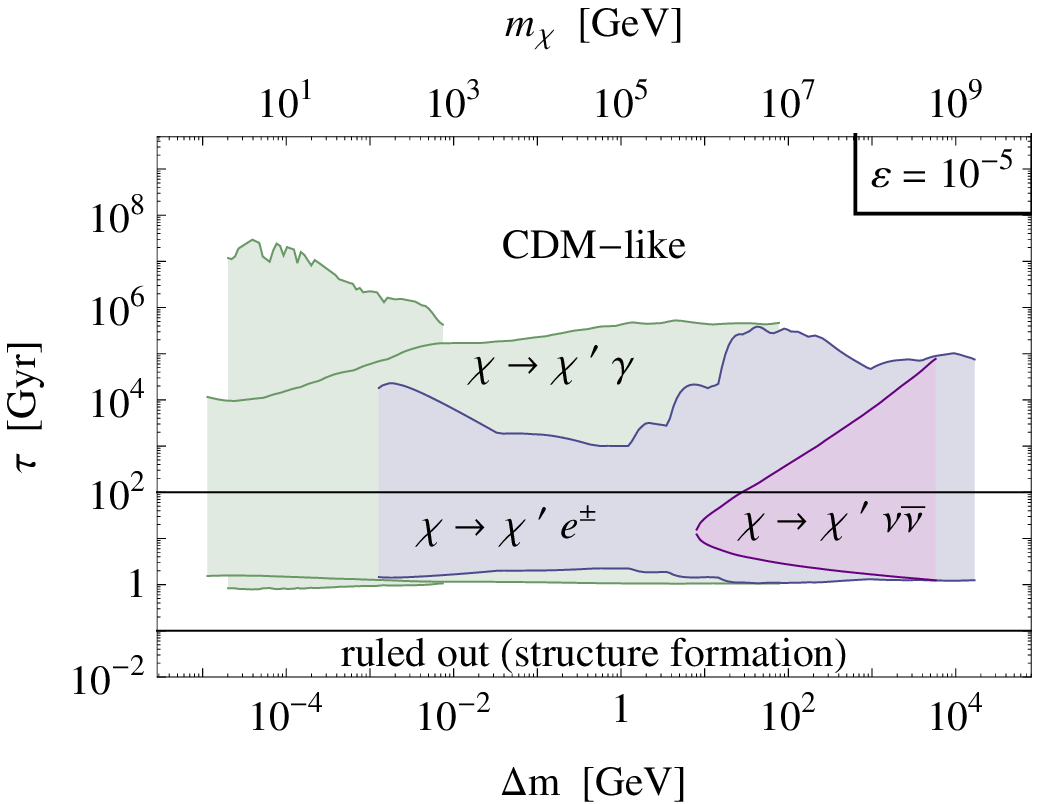}
\caption{Constraints on DM lifetime $\t$ vs mass splitting $\D m$ and
  mass $m_\x$, for $\ve = 10^{-5}$ (recoil velocity $v_{\x'} = 5 \:
  \rm{km/s}$), for the decay channels $\x \rightarrow \x' + \g$
  (green), $\x \rightarrow \x' + e^- + e^+$ (blue) and $\x \rightarrow
  \x' + \n + \bar{\n}$ (purple). Color-shaded regions are excluded by
  the observed radiation backgrounds (Ref.~\cite{Yuksel:2007dr} and
  this work). The decays may have an observable effect on galactic
  halo structure if they occur at times $0.1 \Gyr < \t < 100 \Gyr$
  (interval between solid lines).
  Longer lifetimes $\t \gtrsim 100 \Gyr$ correspond to the standard
  CDM scenario, while decays occurring at $\t \lesssim 0.1 \Gyr$ (but
  after the matter-radiation equality) disrupt the galaxy formation
  significantly and are excluded~\cite{Peter:2010au,Peter:2010jy}.}
\label{TauLimits1}
\end{figure}
\begin{figure}[ht]
\centering
\includegraphics[width=\linewidth]{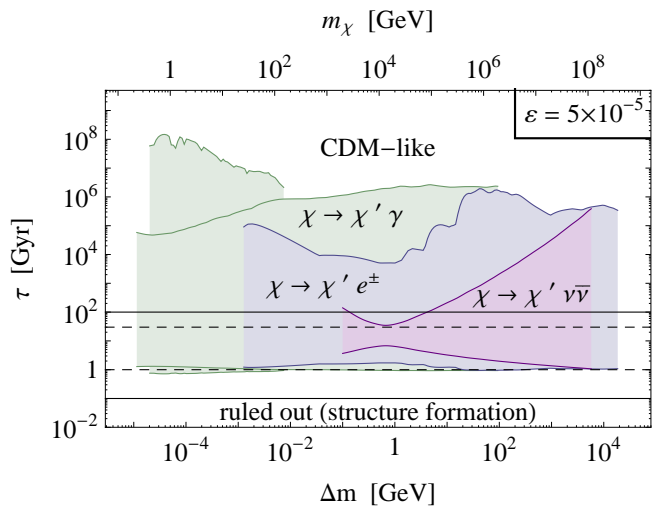}
\caption{Same as in Fig.~\ref{TauLimits1}, but for $\ve= 5 \times
  10^{-5}$, or recoil velocity $v_{\x'} \simeq 15 \: \rm{km/s}$,
  suggested in \cite{Abdelqader:2008wa} as suitable for resolving the
  missing-satellite problem. This would require a dark-matter lifetime
  within the narrower interval $1 \Gyr < \t < 30 \Gyr$ shown (dashed
  lines).}
\label{TauLimits2}
\end{figure}
\begin{figure}[ht]
\centering
\includegraphics[width=\linewidth]{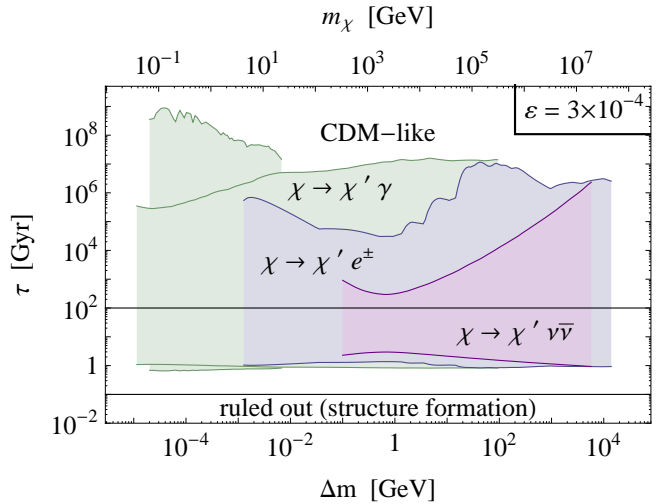}
\caption{Same as in Fig.~\ref{TauLimits1}, but for $\ve= 3 \times
  10^{-4}$, or recoil velocity $v_{\x'} \simeq 90 \: \rm{km/s}$.}
\label{TauLimits3}
\end{figure}

Peter et al.~\cite{Peter:2010au,Peter:2010jy} performed a more
extensive investigation of the $\t-\ve$ parameter space, by means of
both semi-analytical calculations and simulations. They demonstrated
the way in which DM decay affects the halo mass-concentration relation
and mass function,
and used measurements of these quantities to constrain the DM decay
parameter space.  They located the allowed and potentially interesting
parameter space roughly within the range
\bea
\ve &=& 3.4 \times 10^{-6} - 3.4 \times 10^{-4}, \label{epsilon values} \\
\t  &=& (0.1-100) \Gyr.  \label{tau values}
\eea
The corresponding recoil velocity is $v_{\x'} = (1-100) \:
\rm{km/s}$. The actual limits on $\ve$ and $\t$ are, of course,
correlated.

In Sec.~\ref{constraints}, we derived constraints on the lifetime
for late decay of DM into a massive daughter and either $e^\pm$ or a
$\n \bar{\n}$ pair.  For a given value of $\ve$, the bounds of
Sec.~\ref{constraints} determine whether these decays can take place
in time to affect halo evolution.

The flux of DM decay products in the Galaxy today is proportional to $1/(m\t
e^{\t_0/\t})$, and the limits in Fig.\ref{mtauVdm} are expressed in
terms of this quantity.
Translating these limits to bounds on $\t$ alone, we see that the
constraints may be satisfied for either (i) very long lifetimes, $\t
\gg \t_0$, for which the current DM decay rate is very small; or (ii) very
short lifetimes, $\t \ll \t_0$, for which the abundance of the unstable
parent in the Universe today is very small (as most decays have
already taken place).  Between these extremes limits lies a band of
excluded lifetimes.
These bands are shown in Figs.~\ref{TauLimits1} - \ref{TauLimits3} for
three values of $\ve$ spanning the range given in Eq.~(\ref{epsilon values}),
$\ve = 10^{-5}, \: 5 \times 10^{-5}$ and $3 \times 10^{-4}$, corresponding
to recoil velocities
$v_{\x'} \simeq 5 \rm{km/s}, \: 15 \rm{km/s}, \; \rm{and} \; 90 \rm{km/s}$
respectively.

For the values of $\ve$ considered, the excluded band spans
several orders of magnitude in lifetime.  These excluded bands become
smaller as $\ve$ is decreased, such that for sufficiently small
epsilon there is no constraint.
Note that while the upper limit of the excluded band may be improved
with more sensitive flux observations, the same is not true for the lower
limit, at $\t \sim 1$ Gyr, since for these lifetimes the factor
$e^{-\t_0/\t}$ dominates.  For this reason, the lower limit of the
excluded regions for the photon and $e^\pm$ decay modes are very similar.

We now compare the bounds on the DM lifetime $\t$ with the interesting
range of values specified in Eq.~(\ref{tau values}).
Figures~\ref{TauLimits1} - \ref{TauLimits3} show that these bounds
significantly constrain, but do not rule out, unstable
dark-matter models whose dominant decay channel produces the SM
particles considered.
For the $\g$ or $e^\pm$ decay modes, lifetimes greater than $\sim 1
\Gyr$ are eliminated, and we thus identify $\t \sim (0.1 - 1) \Gyr$ as
the range of allowed lifetimes for which decays may affect structure
formation.  For decay modes to neutrinos, however, the allowed
lifetimes span the entire interval of Eq.~(\ref{tau values}) (at least
for some masses).
Of particular interest, perhaps, is the case of 
weakly interacting massive particle (WIMP) dark matter of
mass $m_\x \sim 10^2 \GeV$, decaying via the $\n\bar{\n}$ channel,
with $\t$ and $\ve$ in accordance with the values suggested in
Ref.~\cite{Abdelqader:2008wa} for resolving the missing satellite
problem.  These decay parameters are indicated by the dashed lines in
Fig.~\ref{TauLimits2}.

\section{Limits on models with late dark-matter decay \label{models}}

A number of particle-physics models have been proposed which predict
or invoke a DM decay mode described by Eq.~(\ref{decay
  channel})~\cite{Finkbeiner:2007kk,Finkbeiner:2008gw,TuckerSmith:2001hy,Boubekeur:2010nt}.
The motivation for these models varies widely, thus the parameter
space of interest does not always overlap with that relevant for
structure formation.  We now describe, in general terms, when and how
the bounds derived in Sec.~\ref{constraints} can constrain decay modes
of the type in Eq.~(\ref{decay channel}).  We then apply these
considerations to particular models found in recent literature.

In Secs.~\ref{constraints} and \ref{SSS} we assumed that all of the
DM was in the form of the heavy, unstable, $\x$ particles prior to decay.
However, this need not be the case.  In fact, many models predict or
require that the relic $\x$ abundance be only a fraction of the total
CDM density.
This may be true even if the decay time is larger than the age of the
Universe, provided that either (i) other processes (e.g. inelastic
scattering) efficiently eliminate the heavy particles ($\chi$) in
favor of the light ones ($\chi'$)~\cite{Finkbeiner:2008gw}, or (ii) the
$\chi$ -- $\chi'$ sector makes only a subdominant contribution to the
total DM density.
The relic $\chi$ abundance is, of course, critical in deriving limits
on specific models.

Allowing the $\x$ abundance to be a free parameter, the four variables
which determine the relevance of constraints inferred from the present
radiation backgrounds are (i) the mass splitting, $\D m$, (ii) the
mass of the DM particle, $m_\x$, or equivalently $\ve = \D m/m_\x$,
(iii) the decay lifetime of the heavy DM state, $\t$, and (iv) the
relic fraction of DM in the form of $\x$ particles before decay
occurs, $f_\x \equiv Y_\x/Y\sub{CDM}$.
In terms of these parameters, the constraints of Sec.~\ref{constraints} can be expressed as
\beq
\t e^{\t_0/\t}  \geqslant (\ve \cdot f_\x) \: \mathcal{F}(\D m)/ \D m.
\label{inequality}
\eeq
where $\mathcal{F}(\D m)$ can be read off Fig.~\ref{mtauVdm} for the
decay channels considered.

\subsection{Fixed $\ve \cdot f_\x$}

Consider constraints on the lifetime for models in which the parameters
$\ve$ and $f_\x$ are specified.
Similar to the analysis of Sec.~\ref{SSS}, Eq.~(\ref{inequality})
implies an excluded band for $\t$, provided that the
fraction of energy released per decay (here quantified by $\ve$)
and/or the fraction of decaying dark matter, $f_\x$, are sufficiently
large. More specifically, in order for Eq.~(\ref{inequality}) to
translate into constraints on $\t$, the parameters $\ve, \: f_\x$ must
satisfy
\beq
\ve \cdot f_\x  \geqslant  \frac{\t_0 e}{\mathcal{F}(\D m)/ \D m}.
\label{ef test}
\eeq
(since $\t e^{\t_0/\t} \geqslant \t_0 e$, for any $\t$).

\begin{figure}[t]
\centering
\includegraphics[width=\linewidth]{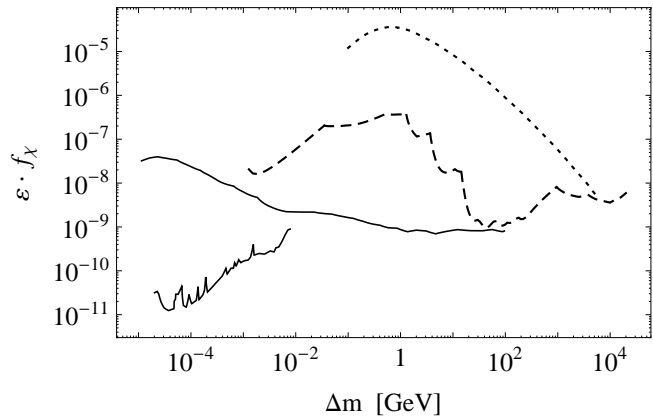}
\caption{
The observed radiation backgrounds can constrain the lifetime of the heavy DM state only in models which lie in the regions above the lines. Models which lie below the lines cannot be constrained by the bounds presented in Sec. \ref{constraints}. As in Fig.~\ref{mtauVdm}, the different curves correspond to the decay channels $\x \rightarrow \x' + \g$ (solid lines), $\x \rightarrow \x' + e^- + e^+$ (dashed line), and $\x \rightarrow \x' + \n + \bar{\n}$ (dotted line).
}
\label{ef}
\end{figure}

In Fig.~\ref{ef} we sketch the regions defined by Eq.~(\ref{ef test}),
for the three decay channels under consideration. 
Models which lie above (below) the curves are constrained (not
constrained) by the bounds derived in Sec.~\ref{constraints} using the
observed radiation backgrounds.
For models that lie on the curves, the constraint (\ref{inequality})
produces a zero-width exclusion band for the lifetime at $\t = \t_0$.
Larger values of $\ve \cdot f_\x$ yield exclusion bands which span a
range of lifetimes around $\t_0$.

The above discussion is relevant to a class of models proposed in
Refs.~\cite{Finkbeiner:2007kk,Finkbeiner:2008gw}, in which DM
possesses an excited state. As a result of an approximate symmetry,
the excited state is separated by a small mass splitting from the
ground state, to which it decays with emission of $e^\pm$ pairs.  For
a particular implementation of these models, with $\D m \approx 2
\MeV$ and $m_\x \approx 500 \GeV$, the kinetic excitations and the
subsequent decays of DM particles inside the Galactic halo can produce
the positrons which give rise to the observed 511~keV annihilation
line in the center of our Galaxy~\cite{Finkbeiner:2007kk}.
For the range of parameters favored by the 511 keV signal, $f_\x
\lesssim 10^{-3}$ and $\ve \simeq 10^{-7} - 10^{-6}$, the scenario
lies comfortably below the $e^\pm$ curve in Fig.~\ref{ef}, and is thus
not constrained by our observational bounds.
However, significantly larger values of $\ve \cdot f_\x$ are possible
within the same generic class of models, for which our bounds would
impose a non-trivial constraint.

\subsection{Fixed $\tau$}

Returning to the constraint of Eq.~(\ref{inequality}), we now consider
models in which $\t$ (rather than $\ve \cdot f_\x$) is narrowly
specified. Equation~(\ref{inequality}) then limits the energy release via
decay, and relic abundance of the unstable state, as per
\beq
\ve \cdot f_\x  \leqslant \frac{\t e^{\t_0/\t}}{\mathcal{F}(\D m)/ \D m}.
\label{ef limit}
\eeq
This constraint is substantial only if
\beq
\t e^{\t_0/\t} \leqslant \mathcal{F}(\D m)/ \D m
\label{tau test}
\eeq
(since by definition $\ve, f_\x \leqslant 1$).
We obtain a nontrivial constraint provided that $ 1 \Gyr \alt \t \alt
10^6 - 10^{12} \Gyr$, where the exact range depends on the mass
splitting and the decay channel considered.
Models with $\t \lesssim 1 \Gyr$ trivially  satisfy the bounds placed
by radiation backgrounds, as do models with very large $\t$, in
accordance with the discussion in Sec.~\ref{SSS}.

The ``degenerate gravitino'' scenario of Ref.~\cite{Boubekeur:2010nt}
falls in the category of models whose energy release may be
constrained by observations.  In this scenario, the lightest supersymmetric particle (LSP) and the next-to-lightest supersymmetric particle (NLSP) have
a small mass splitting, $\D m \sim 10^{-2} \GeV$, such that the
slightly heavier NLSP can decay into the LSP with emission of a
photon, with a lifetime comparable to the age of the Universe.
These parameters satisfy the inequality (\ref{tau test}) and thus the
observational bounds represent a non-trivial constraint on the model.
Specifically, the photon bounds~\cite{Yuksel:2007dr} constrain the
relic NLSP abundance before decay in a mixed (NLSP + LSP) DM scenario,
as explicitly shown in Ref.~\cite{Boubekeur:2010nt}.

As an example of models which trivially evade the constraint (\ref{ef
  limit}), we mention here the inelastic DM
scenarios~\cite{TuckerSmith:2001hy}. These models aim to reconcile the
results of DAMA and CDMS direct detection experiments, by evoking
small mass splittings $\D m \sim 100 \keV$.  
However, since the decays
occur very early (before recombination) this scenario cannot be
constrained by the present radiation backgrounds.  Substituting the
lifetime of the heavier state $\t \sim (10^2 - 10^3) \yr$ into
Eq.~(\ref{tau test}) of course trivially verifies this fact.


\section{Conclusions}

The decay lifetime is a fundamental property of dark matter.
Knowledge of this parameter may provide an essential clue for
unravelling the particle nature of dark matter.  In the standard CDM
scenario, dark matter particle interactions play no role in the
evolution of the Universe subsequent to DM freeze-out, which limits
our ability to probe DM particle interactions via cosmological
observations.  However, there is a large class of DM models in which
DM particle interactions are in fact invoked to explain cosmological
observations.  One such hypothesis is that DM decay may provide a
possible solution to the small scale structure problems of standard
CDM.

We have considered a class of DM models in which DM decays at late
times to a nearly degenerate daughter plus relativistic particles, $\x
\rightarrow \x' + l$.  In this scenario, the relativistic daughters
carry only a small fraction, $\ve = \D m/m_\x$, of the parent DM mass
and thus decays do not alter the DM energy density.  However, the
energy acquired by the daughter $\x'$ effectively heats the halo, and
has been proposed as a means of modifying halo structure in such a way
as to bring CDM predictions into better agreement with observation.

We focused on the scenario in which the relativistic daughters consist
of Standard Model particles, and derived constraints on the cases in
which those daughter particles are either $e^\pm$ or neutrinos.  In
Fig.~\ref{mtauVdm} we summarize these constraints and compare them with
the analogous photon bounds obtained in Ref.~\cite{Yuksel:2007dr}.
Since photons provide the most stringent bounds and neutrinos the
least stringent, these results, taken together, delineate the range of
lifetime bounds for these models.

Our constraints on the $e^\pm$ decay mode lie between those for photons
and neutrinos, though we note that there are some masses for which the
$e^\pm$ constraints become comparable to those for photons.  Note
that many models that allow the process $\x \rightarrow
\x'\bar{\nu}\nu$ will also allow $\x \rightarrow \x'e^+e^-$, for which
stronger bounds apply.

For the $e^\pm$ (and $\g$) decay modes, the lifetime must either be
much greater than the age of the Universe (for which decay cannot
affect structure formation) or smaller than $\sim 1 \Gyr$.  Though
this eliminates some of the parameter space of interest in structure
evolution, a significant portion remains open.  For the $\bar{\nu} \nu$
decay modes, a much larger range of lifetimes is permitted.

Models which are not necessarily motivated by structure formation
considerations, but nonetheless entail decay between particle states
of similar mass, may also be constrained by the bounds presented
here. This is possible, provided that the relic density of the heavier
state from the early Universe is not too small, and/or the decay
lifetime is larger than $1 \Gyr$.
The constraints and the analysis presented here can provide generic
useful guidelines for constructing models which feature decay modes
between closely degenerate states.

We note that our limits in Figs.~\ref{mtauVdm} - \ref{mtauVdm nunubar}
are strictly applicable only for $\ve \ll 1$ (so that the
approximation $E_l \simeq \D m$ is valid).  However, for $\ve=1$, our
monoenergetic constraints correspond to those for the decays
$\x \rightarrow e^+ e^-$ and $\x \rightarrow \bar{\nu}\nu$ (i.e.,
where there is no $\x'$).  For the $\bar{\nu}\nu$ case this simply
reproduces the results of Ref.~\cite{PalomaresRuiz:2007ry}.
For the $e^\pm$ case, our bounds inferred using inverse Compton
scattering and bremsstrahlung emission, synchrotron radiation, and the
positron flux, reproduce the ones derived in
Ref.~\cite{Cirelli:2009dv}, using the same response functions employed
here.  However, the monoenergetic limits of Fig.~\ref{mtauVdm e+e-}
inferred from the in-flight and at-rest annihilations represent strong
new constraints of general applicability for the decay mode $\x
\rightarrow e^+ e^-$, in the interval
$2 \MeV \lesssim m\ssub{DM} \lesssim 1 \GeV$.


\section*{Acknowledgements}
NFB and KP were supported, in part, by the Australian Research
Council, and AJG by the Commonwealth of Australia.  We thank F. Melia,
J. Beacom and H. Y\"{u}ksel for useful discussions.

\bibliographystyle{h-physrev5}
\bibliography{References}

\end{document}

%% file: mymacros.tex
\newcommand{\nc}{\newcommand}
\nc{\sen}[0]{\selectlanguage{english}}
\nc{\sgr}[0]{\selectlanguage{greek}}
\def\cal{\mathcal}
\def\rm{\mathrm}

\def\it{\textit}
\def\sl{\textsl}
\def\bs{\bigskip}
\def\ms{\medskip}
\def\sms{\smallskip}
\nc{\bea}{\begin{eqnarray}}
\nc{\eea}{\end{eqnarray}}
\nc{\beq}{\begin{equation}}
\nc{\eeq}{\end{equation}}
\nc{\bit}{\begin{itemize}}
\nc{\eit}{\end{itemize}}
\nc{\benu}{\begin{enumerate}}
\nc{\eenu}{\end{enumerate}}
\nc{\bdes}{\begin{description}}
\nc{\edes}{\end{description}}

\nc{\nn}{\nonumber}

\nc{\sub}[1]{_{\rm{#1}}}
\nc{\ssub}[1]{_{_\rm{#1}}}
\nc{\super}[1]{^{\rm{#1}}}
\nc{\ssuper}[1]{^{^\rm{#1}}}

\nc{\slashed}[1]{{#1}\hspace{-2mm}/}

\nc{\pare}[1]{\left( #1 \right)}
\nc{\sqpare}[1]{\left[ #1 \right]}
\nc{\ang}[1]{\langle #1 \rangle}
\nc{\abs}[1]{\left| #1 \right|}

\def\g5{\gamma_{5}}

\def \eV{\: \rm{eV}}
\def\keV{\: \rm{keV}}
\def\MeV{\: \rm{MeV}}
\def\GeV{\: \rm{GeV}}
\def\TeV{\: \rm{TeV}}

\def\erg{\: \rm{erg}}

\def \cm{\: \rm{cm}}
\def \km{\: \rm{km}}
\def \pc{\: \rm{pc}}
\def\kpc{\: \rm{kpc}}
\def\Mpc{\: \rm{Mpc}}

\def\sr{\: \rm{sr}}

\def \sec{\: \rm{s}}
\def  \yr{\: \rm{yr}}
\def \Gyr{\: \rm{Gyr}}

\def\a{\alpha}
\def\b{\beta}
\def\g{\gamma}
\def\d{\delta}
\def\e{\epsilon}
\def\z{\zeta}
\def\h{\eta}
\def\th{\theta}
\def\i{\iota}
\def\k{\kappa}
\def\l{\lambda}
\def\m{\mu}
\def\n{\nu}
\def\ks{\xi}
\def\om{o}
\def\p{\pi}
\def\r{\rho}
\def\s{\sigma}
\def\t{\tau}
\def\y{\upsilon}
\def\f{\phi}
\def\x{\chi}
\def\ps{\psi}
\def\w{\omega}

\def\ve{\varepsilon}
\def\vr{\varrho}
\def\vs{\varsigma}
\def\vf{\varphi}

\def\G{\Gamma}
\def\D{\Delta}
\def\Th{\Theta}
\def\L{\Lambda}
\def\Ks{\Xi}
\def\P{\Pi}
\def\S{\Sigma}
\def\Y{\Upsilon}
\def\F{\Phi}
\def\Ps{\Psi}
\def\W{\Omega}
